 \documentstyle[eqsecnum,aps,multicol]{revtex}

\title{Controlled Wavefunction Mixing in a 
Strongly Coupled One-Dimensional Wire}
\author{K.~J. Thomas, J.~T. Nicholls,
M.~Y. Simmons, W.~R. Tribe, A. G. Davies, and M. Pepper}
\address{Cavendish Laboratory, Madingley Road,
Cambridge CB3 0HE, United Kingdom}
\date{\today}
\begin{document}
\maketitle

\begin{abstract}
We investigate the transport properties of
two strongly coupled ballistic one-dimensional (1D) wires,
defined by a split-gate structure deposited
on a GaAs/AlGaAs double quantum well.
Matching the widths and electron densities
of the two wires brings them into resonance,
forming symmetric and antisymmetric 1D subbands
separated by energy gaps that are measured
to be larger than their two-dimensional counterpart.
Applying a magnetic field parallel to the
wire axes enhances wavefunction mixing at low fields,
whereas at high fields the wires become completely decoupled.
\end{abstract}

% \begin{multicols}{2}

The energy levels of a two-level system described by a Hamiltonian $H(a)$
may cross at some particular value of the parameter $a$.
A perturbation applied to the system will couple the two levels,
which will repel each other in accordance
with the ``no-crossing'' theorem\cite{merz70}.
For example, a pair of near-degenerate Stark levels 
in the spectrum of a Rydberg atom in an electric field
can be considered to be such a two-level system\cite{rubb81}.
In this paper we show that similar behavior can be 
observed in a mesoscopic system
when the energy levels of two ballistic 
one-dimensional (1D) wires are brought together.
When the wires have equal width and density,
the symmetry of the 1D wavefunctions allow
the inter-wire interaction (the perturbation) only
to couple subbands of the same index,
giving rise to symmetric and antisymmetric wavefunctions.
The gate voltages controlling the device play
the role of the parameter $a$,
and an in-plane magnetic field
(either parallel $B_{\parallel}$ or perpendicular $B_{\perp}$ to the wire axes)
can tune the inter-wire interaction.

The transport properties of a single ballistic 1D constriction,
defined by a split-gate deposited over a high mobility
two-dimensional electron gas (2DEG), are well known\cite{thomas98b}.
As the voltage applied to the split-gate ($V_{sg}$)
is made more negative the 1D subbands are depopulated,
and the conductance drops by $2e^2/h$ as each subband
passes through the chemical potential.
By fabricating\cite{castle98} a modified split-gate
over a double quantum well sample,
the two ballistic 1D constrictions ($2\times$1D) can be brought together,
one on top of the other (see Fig.~\ref{f:1} inset).
The wafer (T210), grown by molecular beam epitaxy,
comprises two 150~\AA-wide GaAs quantum wells separated by a 25~\AA\
Al$_{0.33}$Ga$_{0.67}$As barrier,
with a center-to-center distance of $d=175$~\AA.
The double quantum well is doped both above and below
using 2000~\AA\ of Si-doped ($1.2 \times 10^{17}$~cm$^{-3}$) AlGaAs,
offset by 600~\AA\ and 800~\AA\ AlGaAs spacer layers, respectively.
The carrier density in each layer is $1.3 \times 10^{11}$~cm$^{-2}$,
with an average mobility of $1.45 \times 10^{6}$~cm$^{2}$/Vs.
The 2D symmetric-antisymmetric (SAS) energy gap
at resonance was measured\cite{davies97} to be $\Delta_{SAS}=1.4$~meV.
The wafer was processed into a Hall bar,
and the gate pattern shown in the Fig.~\ref{f:1} inset
was defined by electron beam lithography.
The split-gate has a width of 1.2~$\mu$m and a length of 0.4~$\mu$m,
and to add further control, a midline gate of width
0.4~$\mu$m was deposited along the center of the split gate.
Our previous measurements\cite{castle98} of  ballistic
$2\times$1D devices show two results:
When the barrier between the two 2DEGs is 300~\AA\ wide,
$\Delta_{SAS} \approx 0$,
and the two 1D wires behave independently with conductances that add.
In a 35~\AA\ barrier sample the coupling is stronger
($\Delta_{SAS} \approx 1$~meV),
and there is mixing of the 1D subbands.
In this paper we have been able to index 
the coupled subbands in a similar sample, 
and show that their behavior in zero and finite in-plane 
magnetic field can be described by a simple model.

The two-terminal conductance $G=dI/dV$ of the parallel wires
was measured at dilution fridge temperatures using standard
techniques\cite{thomas95}.
Conductance traces $G(V_{sg})$
of the $2\times$1D device are shown in Fig.~\ref{f:1},
as the midline voltage $V_{mid}$ was changed
from 0.16 to -1.0~V in 40~mV steps.
After correcting for a series resistance, $R_s=550~\Omega$,
the plateaus in the conductance are quantized in units of $2e^2/h$.
$V_{sg}$ defines both the upper ($u$) and lower ($l$) wires;
however, as $V_{sg}$ becomes more negative measurements show\cite{castle98}
that the subbands in the $l$ wire are depopulated first.
$V_{mid}$ has the opposite effect;
as it is made more negative the $u$ wire is depopulated first.
Therefore, by varying $V_{sg}$ and $V_{mid}$,
the fraction of the current
passing through the $u$ and $l$ wires can be tuned.
The clean conductance steps on
the left-hand and right-hand sides of Fig.~\ref{f:1}
originate from just the $u$ and $l$ wires, respectively.
The more complicated structure in between
occurs where 1D subbands of the two wires become nearly degenerate.

To understand the conductance characteristics better,
they are differentiated\cite{castle98} with respect to $V_{sg}$
to generate the grey scale plot\cite{cool} in Fig.~\ref{f:2}(a).
The dark lines denote the transconductance 
($dG/dV_{sg}$) maxima between plateaus,
indicating where a 1D subband edge passes
through the chemical potential.
The Fig.~\ref{f:2}(a) inset
shows the subband positions redrawn
with new axes ($a$, $b$) and indexed by $n$ and $n^{\ast}$,
which are (to be shown later) the bonding and
antibonding states formed by mixing the $n^{th}$
wavefunctions of the $u$ and $l$ wires.

A magnetic field $B$ applied parallel to the two 2DEGs
shifts the Fermi circles (of radii $k_{F,u}$ and $k_{F,l}$)
with respect to each other by $\Delta k=e B d/\hbar$\cite{eis91b}.
The Fermi circles are completely decoupled when
$\Delta k > k_{F,u} + k_{F,l}$,
which for the carrier densities in our sample occurs for $B > 7$~T.
Figure~\ref{f:2}(b) shows $dG/dV_{sg}$
in an in-plane field $B_{\perp} =8$~T.
A similar plot is obtained for $B_{\parallel} =8$~T,
and for both field directions the characteristics
are similar to those observed\cite{castle98}
when the $u$ and $l$ wires are decoupled in a 300~\AA\ barrier sample.
The subbands in Fig.~\ref{f:2}(b) are indexed with integers $p$ and $q$,
corresponding to the quantization of the conductance (in units of $2e^2/h$)
in the $u$ and $l$ wires.
In Fig.~\ref{f:2}(b) decoupled subbands of the same
index cross without mixing at $a=0$,
the same line along which there are anticrossings
between the $n$ and $n^{\ast}$ subbands at $B=0$ (Fig.~\ref{f:2}(a)).
In addition to this decoupling at $B=8$~T,
there is also evidence of spin-splitting of the 1D subbands.

The potential in a single 1D constriction
is well described by a saddle-point\cite{fertig87,lmm92},
where there is parabolic confinement
in the $y$-direction and a parabolic barrier
to free motion in the $x$-direction.
The conductance (in units of $2e^2/h$)
is determined by the number of occupied 1D subbands at the saddle-point.
For the $2\times$1D device we assume that the two saddle-points
are aligned at the same ($x,y$) position,
and the transport properties are determined
by the mixing of the highest occupied wavefunctions,
$\Psi_j$ (where $j =u$ or $l$), at this point.
If separable, the unperturbed wavefunction
in each wire is given by
\begin{equation}
\Psi_j(x,y,z) = e^{ik_x x}\  \phi_{j,m}(y)\  Z_{j,m}(z),
\label{e:sep}
\end{equation}
where $m=p$ for $j=u$, and $m=q$ for $j=l$.
$\phi_{j,m}(y)$ is the $m^{th}$ 1D subband wavefunction,
and $Z_{j,m}(z)$ is the ground state 2D wavefunction in the quantum well.
$Z_{j,m}(z)$ is a function of the carrier density at the saddle-point,
and hence it depends on the 1D subband index $m$.
For $k_x=0$, there is peak in $dG/dV_{sg}$,
and the matrix element that determines the mixing
between the $p$ and $q$ subbands of the $u$ and $l$ wires is
\begin{equation}
\langle \Psi_{u} | V(z) | \Psi_{l} \rangle  =
\int \phi_{u,p}(y) \phi^{\ast}_{l,q}(y) dy \int Z_{u,p}(z) V(z) Z_{l,q}^{\ast}(z)dz,
\label{e:M}
\end{equation}
where $V(z)$ is the conduction band profile
of the double quantum well structure.
If the integral
\begin{equation}
I= \int \phi_{u,p}(y)\  \phi_{l,q}^{\ast}(y) \ \ dy
\label{e:I}
\end{equation}
is zero, then $ \langle \Psi_{u} | V(z) | \Psi_{l} \rangle = 0$,
and the $p$ and $q$ 1D subbands cross without mixing.
$I$ is determined by both the alignment of the wires and the
symmetries of the two 1D wavefunctions.
If misaligned, then in general $I \neq 0$,
and there will be complicated mixing between all subbands of the two wires;
this is not observed in the $B=0$ results shown in Fig.~\ref{f:2}(a),
supporting the assumption that the
two saddle-points are approximately aligned.
For aligned wires, $I = \delta_{p,q}$, if the wires have the same width.
If the wires have different widths
$I \neq 0$ only when $p$ and $q$ are either both odd or even.
Figure~\ref{f:2}(a) does not show this odd-even 
mixing expected for wires of very different widths,
and so we believe that the aligned wires have 
approximately equal widths\cite{strict}.
In which case the $y$ and $z$ components of the unperturbed wavefunctions
mix when $p=q=n$ to give:
% two different wavefunctions:
\begin{equation}
F_n(y,z) = [\phi_{l,n}(y) Z_{l,n}(z) + \beta \phi_{u,n}(y) Z_{u,n}(z)]/\sqrt{1+\beta^2},
\label{e:s}
\end{equation}
\begin{equation}
F_{n^{\ast}}(y,z) = [\beta \phi_{l,n}(y) Z_{l,n}(z) -
\phi_{u,n}(y) Z_{u,n}(z) ]/\sqrt{1+\beta^2}.
\label{e:as}
\end{equation}
The number $\beta$ describes
how much of the wavefunction is in each wire
and is determined by the position on the $a$-axis,
which is itself a linear combination of $V_{sg}$ and $V_{mid}$.
For example,
the state labeled $n=1$ in Fig.~\ref{f:2}(a)
has the wavefunction $F_1(y)$.
When $a$ is large and negative, $\beta \rightarrow \infty$,
and conduction is through the $u$ wire.
When $a$ is large and positive, $\beta \rightarrow 0$,
and conduction proceeds through the $l$ wire.
At $a=0$ the 1D levels are at resonance ($\beta=1$)
and the integral
\begin{equation}
\int \  Z_{u,n}(z) V(z) Z_{l,n}^{\ast}(z)\ dz = \Delta^{n}_{SAS}/2,
\label{e:J}
\end{equation}
defines the energy gap $\Delta^{n}_{SAS}$ between the
symmetric and antisymmetric 1D subbands at their anticrossing.
Going along the $b$-axis increases the number of occupied 1D subbands.
For example in Fig.~\ref{f:2}(a),
at $a=0$  the symmetric and antisymmetric subbands are populated in the
order $1, 2, 1^{\ast}, 3, 2^{\ast}~\ldots$.
For this reason the grey scale plots
can be thought of as ``energy'' diagrams $E_b(a)$.

At $B=0$ the integral $I=\delta_{p,q}$,
and there is only mixing between subbands of the same index.
Using an in-plane field of $B=2$~T,
we can alter $I$ so that there is also
mixing between subbands of different index;
though, unlike the high $B$ case,
the energy diagram $E_b(a)$ is sensitive
to the direction of the in-plane field.
When $B_{\perp}=2$~T, see Fig.~\ref{f:3}(a),
$E_b(a)$ is similar to that obtained at $B=0$ (Fig.~\ref{f:2}(a)),
and there is no evidence of new anticrossings.
In contrast, for $B_{\parallel}=2$~T, see Fig.~\ref{f:3}(b),
there is strong mixing that
gives rise to complicated crossings and anticrossings.
In a $B_{\parallel}$ field the electron gains momentum
$\Delta k_y = eB_{\parallel} d/ \hbar$
as it tunnels the distance $d$ in the $z$-direction,
and Eq.~\ref{e:I} becomes
\begin{equation}
I (B_{\parallel}) = \int e^{-i\Delta k_y y} \phi_{u,p}(y) \phi_{l,q}^{\ast}(y) \  dy \\
= \int \tilde{\phi}_{u,p}(k_y-\Delta k_y) \tilde{\phi}^{\ast}_{l,q}(k_y) \ dk_y  \\
=\int \tilde{\phi}_{u,p}(k_y- eB_{\parallel}d/\hbar) \tilde{\phi}^{\ast}_{l,q}(k_y) \ dk_y,
\label{e:mix}
\end{equation}
where $\tilde{\phi}_{j,p}(k_y)$ is the Fourier transform of $\phi_{j,p}(y)$.
The wavefunctions of the two wires are shifted in $k_y$-space
and in general $I (B_{\parallel}) \neq 0$.
The $E_b(a)$ diagram at $B_{\parallel}=1$~T (not shown)
exhibits completely different
crossings and anticrossings to those 
at $B_{\parallel}=2$~T (Fig.~\ref{f:3}(b)),
reflecting the changes in $I (B_{\parallel})$.
For the other in-plane direction, $B_{\perp}$,
there is no change in $k_y$ and the observed $E_b(a)$ diagrams
are nearly identical at low fields, $B_{\perp} \ll 7$~T.

To calibrate the $E_b(a)$ diagrams at $B=0$
we have performed DC source-drain voltage measurements
using a spectroscopic technique
developed to study single wires\cite{thomas95}.
The energy spacing between two subbands is given by
the applied DC voltage $V_{sd}$ that causes
their split transconductance ($dG/dV_{sg}$) peaks to cross.
We have used the same technique
to measure $\Delta^n_{SAS}$, the energy gap
at the $n$-$n^{\ast}$ anticrossings along the $b$-axis at $a=0$.
Figure~\ref{f:4}(a) shows the $dG/dV_{sg}$ characteristics
along the $b$-axis at resonance for
$V_{sd}=0, 0.2, 0.4$ and 0.6~mV;
$V_{sd}$ doubles the number of $dG/dV_{sg}$ peaks,
as shown by dashed lines for $n=2$ and 3.
To simplify the picture,
Fig.~\ref{f:4}(b) shows only the position of the right-moving $n$ peaks
and left-moving $n^{\ast}$ peaks as a function of $V_{sd}$.
$\Delta^n_{SAS}$ is the voltage across the $2\times$1D device
at which adjacent $n$ and $n^{\ast}$ peaks cross,
and for this reason the crossing voltages measured
from Fig.~\ref{f:4}(b) are corrected
for series resistance effects\cite{thomas95}.
The corrected values are $\Delta^n_{SAS} =2.6 \pm 0.4$,
$2.4 \pm 0.3$, $2.1 \pm 0.3$ and $1.6 \pm 0.3$~meV for
$n=1$, 2, 3, and 4, respectively.
As $n$ increases the Coulomb repulsion associated with the
higher carrier density in the vicinity of the constriction
pushes the wavefunctions further apart in the $z$-direction,
decreasing the SAS gap towards the 2D limit,
$\Delta_{SAS}=1.4$~meV, of the as-grown sample.

We note that the matrix element in Eq.~\ref{e:M}
is the same as that investigated\cite{mori95}
in 1D resonant tunneling diodes,
where the tunneling current
is proportional to $| \langle \Psi_{u} | V(z) | \Psi_{l} \rangle |^2$.
In these studies the anisotropy in the two
in-plane field directions was observed,
but modeling was required to decide
which tunneling transition $n \rightarrow m$ is associated
with which structure in the $I$-$V$ characteristics.
For the case of strong coupling considered here,
indexing is much easier because the mixing between the $u$ and
$l$ subbands is clearly seen in the $E_b(a)$ diagrams,
and $p$ and $q$ are determined from the quantized conductance.
Other studies\cite{salis98b} have investigated $2\times$1D
wires defined over parabolic quantum wells,
but because of the large well width,
these systems show strong
diamagnetic shifts in an in-plane $B$ field
which complicates the depopulation of the subbands.

In conclusion, we have observed the
bonding and antibonding subbands of coupled 1D wires.
At resonance, the wires are aligned and have approximately equal widths,
and the gap $\Delta^n_{SAS}$ at the anticrossing has been measured.
Using a simple single-particle model we can understand
the overall behavior of the coupled 1D subbands;
there is no evidence for subband locking\cite{sun94},
where 1D subbands are attracted towards each other.
The coupled 1D devices are of interest in
furthering studies of the {\em 0.7 structure} -
this is a reproducible conductance feature at $0.7(2e^2/h)$ 
that is observed\cite{thomas98b,thomas96a} in a clean single 1D wire,
in addition to the usual quantized conductance plateaus.
In a strong parallel magnetic field the 0.7 structure 
evolves into a spin-split plateau at $e^2/h$, 
suggesting that the zero-field structure at $0.7(2e^2/h)$
arises from many-body interactions where spin is an important ingredient.
There are indications of a 0.7 structure in Fig.~\ref{f:1}.
% Other many-body effects could be observed
% in the quantum Hall regime,
% due to the interplay of $\Delta^n_{SAS}$ with the Landau gap
% of the edge states passing through the
% $2 \times$1D constriction.

We thank the Engineering and Physical Sciences Research Council (UK) for
supporting this work, and JTN acknowledges an Advanced EPSRC Fellowship.
We thank Dr. N. J. Appleyard for useful discussions.

% \end{multicols}

%\newpage

\begin{figure}
\caption{Conductance characteristics $G(V_{sg})$ at 1.65~K,
where from left to right $V_{mid}$ is decreased
from 0.16 to -1.0~V in steps of 40~mV.
In the region to the left (right), only the upper (lower) wire is occupied.
Inset: Schematic plan and side view of the device.
The $2\times$1D electron gases are formed 2800~\AA\ below the sample surface;
they lie parallel to the $x$-axis (out of the page),
coupled in the $z$-direction, with 
lateral confinement in the $y$-direction
provided by the split-gate.}
\label{f:1}
\end{figure}

\begin{figure}
\caption{Transconductance ($dG/dV_{sg}$) data at 60~mK presented as a grey scale plot
at (a) $B=0$, and (b) $B_{\perp}=8$~T.
The insets show the indexed subbands plotted with new axes ($a,b$).}
\label{f:2}
\end{figure}

\begin{figure}
\caption{Grey scale plots of $dG/dV_{sg}$
as a function of $V_{mid}$ and $V_{sg}$ at
(a) $B_{\perp}=2$~T, and (b) $B_{\parallel}=2$~T.}
\label{f:3}
\end{figure}

\begin{figure}
\caption{(a) Traces of $dG/dV_{sg}$ on
resonance ($a=0$) at $V_{sd}=0, 0.2, 0.4$ and 0.6~mV,
obtained along the $b$-axis,
a linear combination of $V_{mid}$ and $V_{sg}$,
which are swept simultaneously
across the ranges given at the top and bottom of the figure.
(b) The positions of the right-moving transconductance
peaks of the $n$ subbands ($\bullet$),
and the left-moving peaks of the $n^{\ast}$ subbands ($\circ$)
as a function of $V_{sd}$.
The solid lines are guides to the eye,
showing the crossing voltages used to calculate $\Delta^n_{SAS}$.}
\label{f:4}
\end{figure}


\begin{thebibliography}{10}

\bibitem{merz70}
E. Merzbacher,  in {\em Quantum Mechanics}, 2nd  ed. (Wiley, NY, 1970), pp.\
  428--429.

\bibitem{rubb81}
J.~R. Rubbmark, M.~M. Kash, M.~G. Littman, and D. Kleppner, Phys.\ Rev.\ A {\bf
  23},  3107  (1981).

\bibitem{thomas98b}
K.~J. Thomas {\it et~al.}, Phys.\ Rev.\ B {\bf 58},  4846  (1998).

\bibitem{castle98}
I.~M. Castleton {\it et~al.}, Physica B {\bf 249-251},  157  (1998).

\bibitem{davies97}
A.~G. Davies {\it et~al.}, Phys.\ Rev.\ B {\bf 54},  R17331  (1997).

\bibitem{thomas95}
K.~J. Thomas {\it et~al.}, Appl.\ Phys.\ Lett. {\bf 67},  109  (1995).

\bibitem{cool}
The data presented are from the same device but taken on different cool downs;
  as a result the gate voltage positions shift slightly. The data in Figs.~2-4
  were obtained at 60~mK, and impurity features are visible in the grey scale
  plots shadowing the subband trajectories.

\bibitem{eis91b}
J.~P. Eisenstein, T.~J. Gramila, L.~N. Pfeiffer, and K.~W. West, Phys.\ Rev.\ B
  {\bf 44},  6511  (1991).

\bibitem{fertig87}
H.~A. Fertig and B.~I. Halperin, Phys.\ Rev.\ B {\bf 36},  7969  (1987).

\bibitem{lmm92}
L. Mart\'{\i}n-Moreno, J.~T. Nicholls, N.~K. Patel, and M. Pepper, J.\ Phys.:
  Cond.\ Matt. {\bf 4},  1323  (1992).

\bibitem{strict}
Strictly speaking, the wires only have equal width at $a=0$, but away from
  resonance the mixing described by $I$ is extremely weak for $p \neq q$.

\bibitem{mori95}
N. Mori, P.~H. Beton, J. Wang, and L. Eaves, Phys.\ Rev.\ B {\bf 51},  1735
  (1995).

\bibitem{salis98b}
G. Salis {\it et~al.},   (1998), {P}roceedings of the Symposium on {\em
  Research in High Magnetic Fields}, Nijmegen, to be published.

\bibitem{sun94}
Y. Sun and G. Kirczenow, Phys.\ Rev.\ Lett. {\bf 72},  674  (1994).

\bibitem{thomas96a}
K.~J. Thomas {\it et~al.}, Phys.\ Rev.\ Lett. {\bf 77},  135  (1996).

\end{thebibliography}
\end{document}